\documentclass{article}
\usepackage{spconf,amsmath,graphicx,multicol,multirow,booktabs}
\usepackage{amssymb}
\usepackage{enumitem}
\setlist{nosep, leftmargin=14pt}

\usepackage{mwe} 

\title{Fully differentiable correlation-driven 2D/3D registration for X-Ray to CT Image Fusion}

\name{Minheng Chen$^{1}$ \qquad Zhirun Zhang$^{1}$ \qquad  Shuheng Gu$^{1}$ \qquad Zhangyang Ge$^{1}$ \qquad Youyong Kong$^{1 2 3 \star}$ \thanks{$^{\star}$Corresponding author. (Email: 
kongyouyong@seu.edu.cn ) }}
\address{$^{1}$ School of Computer Science and Engineering, Southeast University, China\\
$^{2}$Jiangsu Provincial Joint International Research Laboratory of Medical \\Information Processing,  Southeast University, China \\
$^{3}$Key Laboratory of New Generation Artificial Intelligence Technology and  Its Interdisciplinary \\Applications (Southeast University), Ministry of Education, China}
\begin{document}
%
\maketitle
\begin{abstract}
Image-based rigid 2D/3D registration is a critical technique 
for fluoroscopic guided surgical interventions. 
In recent years, some learning-based fully differentiable methods have produced beneficial outcomes while the process of feature extraction and gradient flow transmission still lack controllability and interpretability. 
To alleviate these problems, in this work, we propose a novel fully differentiable correlation-driven network using a dual-branch CNN-transformer encoder which enables the network to extract and separate low-frequency global features from high-frequency local features. 
A correlation-driven loss is further proposed for low-frequency feature and high-frequency feature decomposition based on embedded information. Besides, a training strategy that learns to approximate a convex-shape similarity function is applied in our work. We test our approach  on a in-house dataset
and show that it outperforms both existing fully differentiable learning-based registration approaches and the conventional optimization-based baseline.
Our code is available at \textit{https://github.com/m1nhengChen/cdreg}.
\end{abstract}
\section{Introduction}
\label{sec:intro}
Image guidance for minimally invasive interventions is generally provided using live fluoroscopic X-ray imaging. The fusion of preoperative Computed Tomography (CT) volume with the live fluoroscopic image enhances the information available during the intervention. Rigid 2D/3D registration computes the pose geometry of 3D objects from intra-operative 2D images, which is essential for  accurate fusion with the fluoroscopic image. Traditionally, optimization-based techniques~\cite{chen2024optimization,op2patch} have been utilized for 2D/3D registration in interventional procedures due to their high accuracy. However, these techniques are sensitive to initialization and content mismatch between X-ray and CT images, resulting in small capture ranges and a tendency to fall into local optima. As a consequence, registration failure may occur when the initial pose is far from the ground truth.

With the advancements in deep learning techniques, the remarkable representation learning capabilities have had a significant influence on 2D/3D registration. In particular, 
researchers have recently proposed a fully differentiable framework~\cite{prost,gao2023fully}. Specifically, this approach employs a CNN encoder to independently embed the target 2D images and the  2D Digitally Reconstructed Radiography (DRR) images generated from 3D objects. By assessing the similarity between these embedded representations in the latent space, the pose is updated accordingly. This technique effectively enhances the capture range of 2D/3D registration and improves the overall registration performance.

However, existing fully differentiable methods have some shortcomings. First, the intricate internal mechanisms of CNNs is difficult to control and interpret. This complexity can lead to inadequate feature extraction of DRR and X-ray images, ultimately compromising the reliability of this approach in clinical applications. Second, existing context-free CNN architectures typically focus on extracting local information within a limited receptive field. Consequently, it becomes challenging to capture global information necessary for achieving highly accurate registration results. As a result, it remains unclear whether the inductive bias of CNNs alone is sufficient to extract the required features for optimal performance in this context.
To overcome the above shortcomings, we propose a more reasonable paradigm.

First, our assumption is that, previous optimization-based methods~\cite{op2patch} use gradient correlation 
as a similarity metric for registration, which means that the goal of this process is to gradually increase the correlation of high-frequency information. At the same time, due to the disparity in imaging effects between DRRs and X-rays, low-frequency information tends to introduce significant interference.
Therefore, we strive to separate low-frequency information that is irrelevant to registration during the feature extraction process.
To achieve this,
\begin{figure*}[htb]

  \centering
  \centerline{\includegraphics[width=\linewidth]{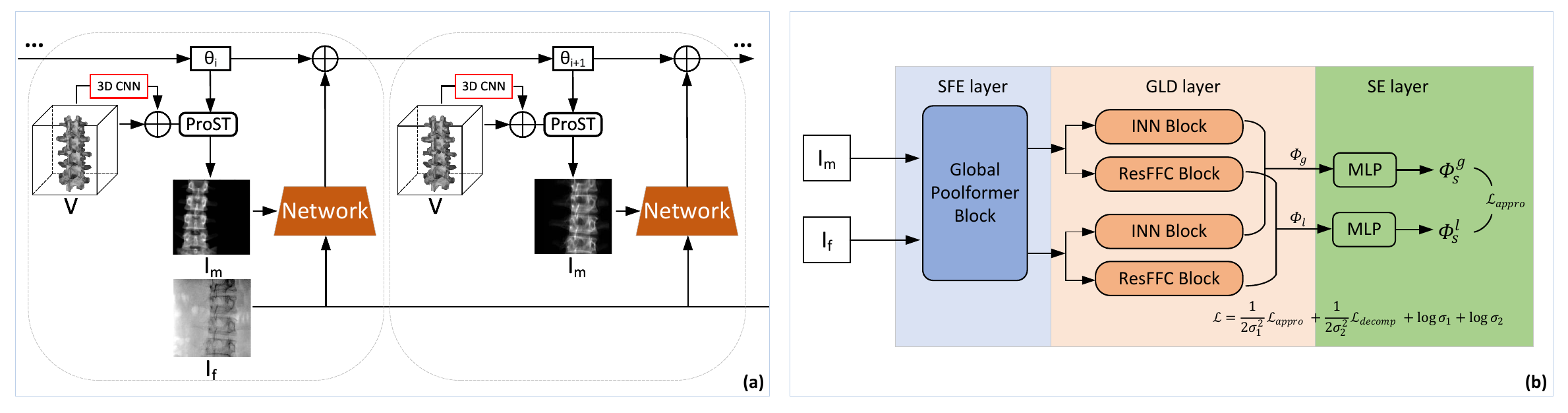}}
  \vspace{-0.5cm}
%
%
\caption{Overall architecture of our Method. (a) It includes the architecture of the proposed framework, which is trained to predict a relative SE(3) transformation that can be applied to an iterative 2D/3D registration. (b) The structure of the encoder  consists of three components: the shallow share feature encoder(SFE), the global-local feature decomposition(GLD) layer and the similarity evaluation (SE) layer.}
\label{fig:overall}
\end{figure*}
we introduce a correlation-driven loss function  that enhances the feature decomposition of DRRs and X-rays.
This method ensures that the high-frequency detailed features are correlated while the low-frequency global features remain uncorrelated, which effectively suppresses redundant information and improves the interpretability and controllability of the process. 

Second, we propose a dual-branch encoder that combines the strengths of CNN and transformer architectures. This novel approach allows us to extract and decompose local and global features, which better reflects the specific and shared features of DRRs and X-rays. Additionally, quantitative performance is reported comparing with representative optimization-based methods and previous fully differentiable methods, using the conventional CMA-ES registration methods as a benchmark. In summary, our main contributions are as follows:

\begin{itemize}
\item  We introduce a dual-branch CNN-Transformer encoder which allows us to extract and decompose local and global features.
\item  We propose a  correlation-driven loss for low-frequency feature and high-frequency feature decomposition.
\item  We adopt a training strategy that aims to learn an approximation of a convex-shaped similarity function. 
\item  Furthermore, we demonstrate superiority of our correla-tion-driven framework on synthetic data.
\end{itemize}
\section{method}
In this section, we will first introduce the workflow of our method and the detailed structure of each module. We will also discuss the training strategy for the proposed method.
\vspace{-0.2cm} 
\subsection{Overview}
\label{sec:over}
The problem of rigid CT to X-ray registration can be formulated
as follows: Given a fixed 2D X-ray image $\mathit{I}_x$  and a moving 3D volume $\textit{V}$ as input, the 2D X-ray to 3D volume registration problem is to seek a mapping function $\mathcal{F}$ to retrieve the pose parameter $\theta \in \mathrm{SE(3)}$ such that the image simulated from the 3D CT is as similar as possible to the acquired image $\mathit{I}_x$: 
\vspace{-0.15cm} 
\begin{equation}\label{eq1}
\begin{split}
\mathcal{F}(\theta) &=arg\mathop{min}_{\theta}\mathcal{S}(\mathit{I}_x,\mathit{I}_m)\\
 &= arg\mathop{min}_{\theta}\mathcal{S}(\mathit{I}_x,\mathit{P}(\theta;\textit{V}))
\end{split}
\end{equation}
where $\mathcal{S}$ represents a similarity metric between the intraoperative fluoroscopic image $\mathit{I}_x$ and the DRR $\mathit{I}_m$. $\mathit{P}(\theta;\textit{V})$ denotes the generation of DRR  from volumetric 3D scene $\textit{V}$ by using a 6 DoF pose $\theta$ and projection operator $\mathit{P}$. 

Since our network is fully differentiable, the registration can be viewed as a gradient-based iterative optimization process. The output of the well-trained network $\phi$ with fixed parameters can be considered as a similarity objective function.
And the $i$-th stage of this iterative alignment can be shown as:
\begin{equation}\label{eq2}
\theta_{i} =\theta_{i-1}-\alpha \frac{\partial\phi(\theta_{i-1})}{\partial\mathit{I}^{i-1}_m}\frac{\partial\mathit{I}^{i-1}_m}{\partial\mathit{P}(\theta_{i-1};\textit{V})}\frac{\partial\mathit{P}(\theta_{i-1};\textit{V})}{\partial\theta_{i-1}}
\end{equation}

\subsection{Network Architecture}
\label{sec:net}
Figure.~\ref{fig:overall}(a) shows the architecture of the proposed framework.
Given an input volume $\textit{V} \in \mathbb{R}^{H\times W\times D}$ , a fixed 2D image $\mathit{I}_x \in \mathbb{R}^{H\times W}$ and an initial pose $\theta_{ini} \in \mathrm{SE(3)}$ , where H, W and D denote the height, width, and depth, respectively. 

Following previous methods~\cite{prost,gao2023fully}, we employ a 3D CNN to learn the residual from $\textit{V}$. The projected moving image $\mathit{I}_m$ is generated by using the ProST~\cite{prost} projection module. And then $\mathit{I}_m$ and $\mathit{I}_x$ will pass through the proposed dual-branch CNN-Transformer encoder. The structure of the encoder is illustrated in Fig.\ref{fig:overall}(b), which consists of three components: the shallow share feature encoder(SFE), the global-local feature decomposition(GLD) layer and the similarity evaluation (SE) layer.
\\
\noindent\textbf{Shallow share feature encoder.} The objective of SFE is to extract shallow features from $\mathit{I}_m$ and $\mathit{I}_x$ individually. To achieve this, we propose a weight-shared Global Poolformer module~\cite{xing2022nestedformer}.  As discussed in ~\cite{yu2022metaformer}, it has been demonstrated that a Poolformer-like structure can outperform recent transformer and MLP-like models. 
The computationally intensive attention module in Transformer is swapped out for a global pooling operation to provide this improved performance.
This module helps in extracting shared shallow spatial features, with a particular focus on capturing global dependencies.

\noindent\textbf{Global-local feature decomposition layer.} The GLD layer aims to extract and decouple global and local features from  the shared features $\Phi^{x}_s$ and $\Phi^{m}_s$. 
We use two branches to extract local and global features respectively. 
Fast Fourier convolution~\cite{chi2020fastfc} is a recently proposed operator that enables the utilization of global context in early layers while providing a receptive field that spans the entire image. 
This makes it particularly well-suited for capturing the global information of an image. 
Thus we employ a residual fast Fourier convolution block~\cite{ffc} for extracting global features.

Invertible Neural Network (INN) is advantageous as it effectively preserves the input information by ensuring that its input and output features generate each other. 
This aligns with our desired outcome of retaining important input information during the local feature extraction process. 
Therefore, we adopt the INN block with affine coupling layers~\cite{zhao2023cddfuse}. 
Let the global feature extraction and local feature extraction be represented by $\mathit{G}(\cdot)$ and $\mathit{L}(\cdot)$ respectively. And this procedure can be formulated as:
\begin{equation}\label{eq3}
\Phi_g =(\mathit{G}(\Phi^{x}_s)-\mathit{G}(\Phi^{m}_s))^2,\Phi_l =(\mathit{L}(\Phi^{x}_s)-\mathit{L}(\Phi^{m}_s))^2
\end{equation}
\noindent\textbf{Similarity evaluation layer.} The function of the SE layer is to estimate image similarity on
local/global features $\Phi_l$ and $\Phi_g$, respectively. Specifically, we employ 2 small MLPs with 5 fully connected layers and ReLU activation
functions in between to learn the global/local similarity estimation. 
And the output of this layer can be represent as $\Phi^s_g$ and $\Phi^s_l$.
\subsection{Training and Testing}
A straightforward way to train the registration network is to use L2 distance to measure the rotation and translation error.  However, this method suffers from the difficulty of balancing the weight of rotation loss and translation loss.
Like previous fully differentiable 2D/3D registration methods~\cite{prost,chen2023embeddedsopi}, we adopt a training strategy named double backward mechanism to learn to approximate the convex shape of geodesic loss, $\mathcal{L}_{geo}(\theta, \theta_t)$, which is the square of the geodesic distance between current pose $\theta$ and target pose $\theta_t$ in $\mathrm{SE(3)}$. This method avoids the impact of absolute loss scale by approximating the gradient of the network to the gradient of geodesic distance.

Specifically, at training stage, the gradient approximation loss $\mathcal{L}$ is:
\begin{equation}
\label{eq4total}
\mathcal{L}_{appro}=\mathcal{L}_{geo}(\frac{\partial\mathcal{L}_{net}}{\partial\theta},\frac{\partial\mathcal{L}_{geo}(\theta, \theta_t)}{\partial\theta})
\end{equation}
\begin{equation}\label{eq5net}
\mathcal{L}_{net}=\langle \sigma(\Phi^s_g), \sigma(\Phi^s_l) \rangle
\end{equation}
where $\mathcal{L}_{net}$ is the loss of the network.
As shown in Eq.~\ref{eq6ncc}, we propose a  correlation-based decomposition loss $\mathcal{L}_{decomp}$ which uses the normalized cross correlation operator $\mathrm{NCC}(\cdot,\cdot)$ to decouple the local/global information. 
\begin{equation}\label{eq6ncc}
\mathcal{L}_{decomp}=\frac{\mathrm{NCC}(\mathit{G}(\Phi^{x}_s), \mathit{G}(\Phi^{m}_s))}{\mathrm{NCC}(\mathit{L}(\Phi^{x}_s),\mathit{L}(\Phi^{m}_s))+\epsilon}
\end{equation}
However, achieving a balance in the weights of multiple losses is still a challenging task, even when employing the time-consuming grid search methods to tune hyperparameters. This makes the gradient flow transmission process difficult to control.  As a result, we opt for a loss function that incorporates uncertain weights~\cite{kendall2018multi} during training: 
\begin{equation}\label{eq7total}
\mathcal{L}=\frac{1}{2\sigma^2_1}\mathcal{L}_{appro}+ \frac{1}{2\sigma^2_2}\mathcal{L}_{decomp} + \log\sigma_1 + \log\sigma_2
\end{equation}
where $\sigma_1$, $\sigma_2$ are learnable variables and $\epsilon$ is hyperparameter.

During inference, the network performs gradient-based optimization over $\theta$ based on the back-propagation gradient flow.
The update of $\theta$ in each iteration follows the description provided in Section-~\ref{sec:over}. 
This optimization process is implemented using PyTorch's Stochastic Gradient Descent (SGD) optimizer.
\vspace{-0.2cm}
\section{experiment}
We evaluate our method on simulated X-ray images 
for a challenging single-view 2D/3D lumbar spine registration scenario.
\subsection{Dataset and Evaluation Metrics}

\noindent\textbf{Dataset.} The dataset consists of 465 CT scans from collaborating institutions. The spines are segmented using an automatic method in~\cite{zhang2024spineclue}. We resample the CT images to isotropic spacing of 1.0 mm and crop or pad evenly along each dimension to obtain  $256\times256\times256$ volumes with the spine ROI approximately in the center. We select 418 scans for training and validation, and 47 scans are used for testing. We simulate X-rays with resolution of 0.798 mm × 0.798 mm, and size of 256 × 256. For testing, we use 500 simulated X-ray images with angles of $N$(0, 20) degrees in three directions, with translation in mm of $N$(0, 30) for in-plane (X and Y) direction and $N$(0, 60) for depth (Z) direction.

\noindent\textbf{Evaluation metrics.} Following the standard in 2D/3D registration~\cite{van2005standardized}, we use the following two evaluation metrics for all our experiments.
\begin{itemize}
    \item \textit{Mean Target Registration Error (mTRE).} This metric computes the mean distance of corresponding landmarks between the warped and the target image.
The mTREs reported in our experiments are the top 50\%, 75\%, and 95\% (in millimeters) of the synthesis images.
     \item \textit{Success Rate (SR).} In addition, we also report the success rate of the registration, which is defined as the percentage of the tested cases with a TRE smaller than 10 mm.
\end{itemize}
\subsection{Implementation Details}
We define the intrinsic parameter of the X-ray simulation environment as a Perlove PLX118F C-Arm, which  has image dimensions of $1024 \times 1024$, isotropic pixel spacing of 0.199 mm/pixel, and a source-to-detector distance of 1012 mm. The images are downsampled to have dimensions of $256\times256$ with a pixel spacing of 0.798 mm/pixel. 

During training iteration i, we randomly sample a pair of pose parameters ($\theta^i$, $\theta^i_t$) with rotations from a normal distribution $N$(0, 10)  in degrees for all three axes, and translations $t_x,t_y,t_z$ from normal distributions $N$(0, 30), $N$(0, 15) and $N$(0, 15)  in millimeters. Additionally, we randomly select a CT volume denoted as $\textit{V}$ and its corresponding segment $V_{seg}$. The target image is generated in real-time using 
$\textit{V}$ and $\theta^i_t$. $V_{seg}$ and $\theta^i$ are used as input to our network. The proposed framework was
implemented with PyTorch on NVIDIA GeForce RTX 3090 GPUs with 24 GB memory. The model wass trained by using a SGD optimizer with a cyclic learning
rate between 10 e-6 and 10 e-4 every 100 steps and
a momentum of 0.9 for 200 k iterations. And a domain randomization strategy proposed in~\cite{grimm2021poserandomization} was used during training.

\begin{table}[h!]
\begin{center}
	\label{table1}
 
	\caption{2D/3D registration performance comparing with the baseline methods. This evaluation includes measurement of the mean Target Registration Error (mTRE) at top 50\%, 75\%, and 95\%, as well as calculating the success rate (SR) of registration.}
 \setlength{\tabcolsep}{0.7mm}{
	\begin{tabular}{c|ccc|c} 
  \toprule[1.2pt]
  \multirow{2}{*}{Method}
  & \multicolumn{3}{|c|}{mTRE(mm)$\downarrow$}& \multirow{2}{*}{SR(\%)$\uparrow$} 
		\\
   \cline{2-4}    & Top 95\% & Top 75\% & Top 50\% & 
  \\
\hline
  Initial & 225.7$\pm$ 89.7 &188.6$\pm$64.5 & 148.0$\pm$47.3&    \\
  +CMA-ES & 98.6$\pm$98.6&55.7$\pm$49.9 & 24.2$\pm$14.7 & 22.0  \\
\hline
ProST &185.7$\pm$ 90.2 &151.2$\pm$ 65.8 &114.3$\pm$46.3 &  \\
  +CMA-ES &38.3$\pm$55.5 &12.8$\pm$19.7 &2.6$\pm$1.3 &  55.6\\
\hline
SOPI &163.6$\pm$78.9 &133.1$\pm$56.9 &101.3$\pm$39.9 &  \\
  +CMA-ES &34.7$\pm$55.8 &9.1 $\pm$13.7&2.2$\pm$1.0 & 58.4 \\
\hline
\textbf{Ours} & 155.7$\pm$69.6& 127.2$\pm$43.8& 95.1$\pm$30.1&  \\
  +CMA-ES &32.0$\pm$51.9 &7.9$\pm$11.9 &2.2$\pm$0.9 & \textbf{61.0} \\
\bottomrule[1.2pt]
	\end{tabular}}
\label{result}	
\end{center}
\end{table}
 
\vspace{-0.8cm}
\subsection{Comparison with Existing Methods}
Similar to the experiments conducted in other articles~\cite{prost,gao2023fully}, our experiments aim to demonstrate a significant increase in the capture range of 2D/3D registration when optimization-based method is applied after utilizing our network. We employ CMA-ES~\cite{hansen2003reducingCMAES} as the optimization-based baseline. 
A detailed introduction to the CMA-ES baseline can be found in~\cite{chen2024optimization}.
And we conduct comparisons with the other fully differentiable learning-based 2D/3D registration methods: ProST~\cite{prost} and SOPI~\cite{chen2023embeddedsopi}. To ensure fairness, we exclude the pose initialization module from SOPI. Additionally, we employ publicly available implementations of these approaches and adhere to the hyperparameter settings outlined in the original paper. 

 

\begin{figure}[htb]

\begin{minipage}[b]{1.0\linewidth}
  \centering
\vspace{-0.2cm}
  \centerline{\includegraphics[width=\linewidth]{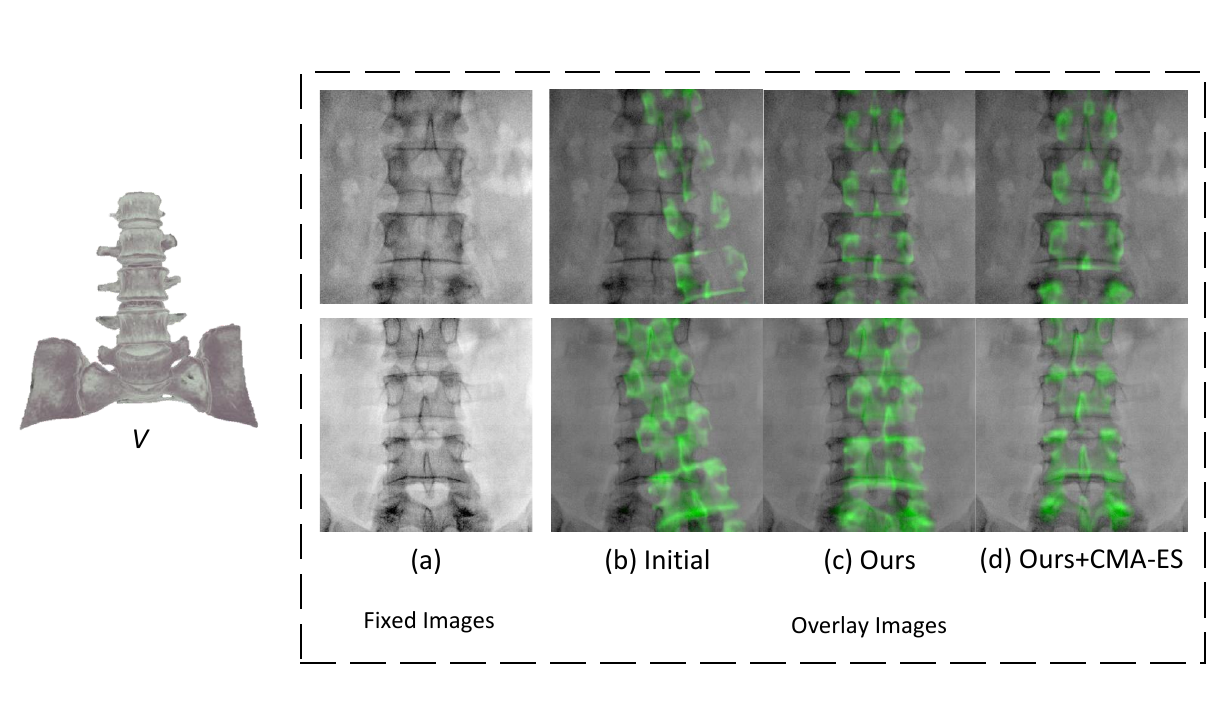}}
  \vspace{-0.6cm}
\end{minipage}
%
%
\caption{Quantitative results  on a test dataset  using our proposed method. Each column in the figures represents: (a) fixed images (b) overlay images of initial pose (c) overlay results after applying the proposed method (d) visualization results after employing the proposed method and CMA-ES. The overlay images are created by superimposing the fixed images with the DRR-derived edges highlighted in green.}
\label{fig:result}
\end{figure}

As shown in Table~\ref{result}, 
our method outperforms existing fully-differentiable methods in terms of the top 50\%, 75\%, and 95\%  of the mTRE, demonstrating superior performance. Additionally, our approach exhibits a higher success rate throughout the experiment, indicating its robustness. Furthermore, it suggests a broader capture range compared to existing methods.
Moreover, we provide several qualitative results of our proposed registration method in Fig.~\ref{fig:result}. The robust performance of this method demonstrates its strong controllability.

\vspace{-0.7cm}
\section{conclusion}
In this work, we propose a novel fully differentiable correlati-on-driven network for 2D/3D registration. 
We aim to tackle the problem of poor interpretability and controllability, as well as the limited capture range in existing end-to-end differentiable methods. The former two problems are addressed by the assumption of gradual increase in correlation of high-frequency
information and the latter solved by introducing a dual-branch CNN-Transformer encoder. The experiments above demonstrate the effectiveness and robustness of our method 
and we believe the fully differentiable correlation-driven method for registration worth further attention and researching.

\section{Compliance with Ethical Standards}
This research study was conducted using human subject data.
The institutional review board at the local institution approved
the acquisition of the data, and written consent was obtained
from the subject.

\section{acknowledgement}
This work was supported in part by Bond-Star Medical Technology Co., Ltd..  
We thank Sheng Zhang, Tonglong Li, Junxian Wu and Ziyue Zhang for their constructive suggestions at several stages of the project.

\bibliographystyle{IEEEbib}
\bibliography{strings,refs}

\end{document}